# Studying seismic activity of the earthquake in India using fractal analysis

Dr. Santanu Nandi

Dept. of Mathematics, School of Advanced Sciences, VIT - AP University, Andhra Pradesh - 522241, India.

Email: santanu282@gmail.com

## ABSTRACT:

Natural disaster strikes at any given moment from seemingly out of nowhere Akin to earthquake that strongly affects human with different magnitudes through the course of time. The main aim of this study is the fractal analysis of seismic activity data of India in the interval from 04-10-2016 to 31-05-2023. This includes analyzing the earthquake magnitudes and their epicenters using fractal statistics, which were studied at different scales to identify patterns in the data through the use of the fractal spectrum. The probabilities of future earthquakes with different magnitudes were estimated using the fractal model.

***Keywords:*** *Fractal Analysis, Multifractal Analysis, Seismic Data Modelling, Earthquake Prediction, Fractal, Dimension, Kolmogorov-Smirnov Test, Anomaly Detection*

## 1. INTRODUCTION:

The study of earthquake dynamics is a critical area of research due to its potential to mitigate natural disaster risks and improve the predictive understanding of seismic events. Traditional methods for analysing earthquake data primarily focus on linear statistical approaches, but these often fail to capture the complex, non-linear nature of seismic activities. In this context, fractal analysis has emerged as a powerful tool for exploring the underlying patterns and self-similarity inherent in earthquake occurrences.

Fractals provide a mathematical framework for describing irregular, fragmented, and self-similar structures found in nature, and are widely applied in geophysical phenomena such as fault systems, earthquake magnitudes, and spatial distributions. The fractal dimension, in particular, is an effective measure of the complexity and spatial distribution of earthquakes, offering insights into how seismic events are clustered in space and time.

In this work, we apply fractal dimension analysis to earthquake data collected across India during the period of October 2016 to May 2023. By analysing the magnitude and spatial coordinates of seismic events, we aim to estimate the fractal dimensions of earthquake distributions, uncovering the degree of clustering and the scale-invariant properties of seismic activity. Specifically, we compute the fractal dimension for earthquake magnitudes using techniques derived from multifractal analysis, as well as the relationship between magnitude and scale.

This study contributes to the understanding of earthquake distribution in India by applying fractal geometry, providing a novel perspective that may improve risk assessments and enhance seismic hazard prediction models in the region.

## 2. LITERATURE REVIEW:

Fractal and multifractal analysis have become prominent tools in understanding the complex patterns inherent in seismic data. **Smalley et al. (1987)** pioneered the use of fractal dimensions to study earthquake clustering, applying this approach to the seismicity of the New Hebrides and demonstrating the potential of fractals in analysing non-linear characteristics in seismic data [1]. Building on these foundational insights, **Kagan (1991)** extended the application of fractal and scale-invariant methods to the distribution of seismic moments, emphasizing how seismic events exhibit consistent patterns across different scales [2].

Early applications of fractal analysis often relied on **mono-fractal models**, which used a single fractal dimension to describe seismic patterns. For instance, **Lapenna et al. (2004)** applied mono-fractal analysis to seismic sequences but acknowledged that this approach oversimplified the data's complexity [3]. The need for more comprehensive models led to the shift toward **multifractal analysis**, which can capture diverse scaling behaviours in seismic data. **Padua et al. (2013)** highlighted the limitations of mono-fractal models in seismic studies and demonstrated the use of multifractal distributions to better fit seismic data from the Philippines, showing improved analysis of local and global variability [4].

Further supporting this approach, **Telesca, Lapenna, and Macchiato (2004)** investigated scaling properties in seismic sequences using both mono- and multifractal techniques, showcasing how multifractal models could reveal hidden structures in the temporal distribution of seismic events [3]. These multifractal models have proven valuable in understanding the variability and clustering of seismic activity, which mono-fractal methods often overlook.

The concept of fractal dimensions extends beyond seismology. **Liebovitch and Shehadeh (2003)** provided a comprehensive introduction to fractal analysis, laying the groundwork for its application in various scientific fields [6]. Additionally, the application of multifractal spectral analysis to non-seismic phenomena, such as financial markets, was explored by **Los and Yalamova (2004)**, who applied multifractal techniques to study the 1987 stock market crash, reinforcing the versatility of these analytical methods [7].

In more recent studies, **Valmores et al. (2014)** applied fractal dimension analysis to psychological data, further showcasing the method's adaptability across disciplines [5]. This cross-disciplinary application underlines the robustness of fractal and multifractal analysis in capturing complex, non-linear patterns.

In summary, while mono-fractal models provided initial insights into seismic patterns, subsequent studies, such as those by **Padua et al. (2013)** and **Telesca et al. (2004)**, demonstrated that multifractal models offer a deeper understanding of seismic data's heterogeneous and scale-dependent behavior. The evolution from mono-fractal to multifractal methods has improved our ability to model and predict seismic phenomena, providing a more nuanced approach to understanding the distribution and variability of earthquake events.

# 3. METHODOLOGY:

## 3.1 Dataset Description:

The earthquake dataset used in this study was sourced from the National Centre for Seismology (NCS), a government institution in India, and covers seismic events in India between October 4, 2016, and May 31, 2023. The dataset provides detailed records of earthquake occurrences.

Web Link - Data Portal | Official website of National Center for Seismology, Ministry of Earth Sciences, Government of India

## 3.2 Data Preprocessing:

The earthquake data was processed to check for missing values and outliers, resulting in a well-cleaned dataset. An additional column for state names was created from the detailed location information in the location column.

## 3.3 Data distribution analysis:

The dataset, consisting of earthquake events recorded from October 2016 to May 2023, showed a wide range of magnitudes from 1.0 to 7.5, with a mean magnitude of 3.8 and a standard deviation of 1.2. The distribution of earthquake magnitudes was visualized using a histogram and Q-Q plot.

Further, normality tests, including the Kolmogorov-Smirnov test, indicated that the magnitude data did not follow a normal distribution (p-value < 0.05). This non-normal distribution aligns with the power-law nature of seismic activity, where smaller earthquakes are more frequent than larger ones. The presence of outliers was noted, particularly high-magnitude events that are rare but significant for analysis.

These characteristics support the application of fractal analysis, as the data exhibits non-uniformity and complex scaling properties across different magnitudes, indicative of fractal behaviour.

## 3.4 Fractal analysis for calculating probability:

Statistical fractal observations are relatively new in the discipline of Statistics. While classical statistics depend on the existence of a mean (μ) and variance (σ2) of a set of normally distributed random observations [N(μ,σ2)], most real-life observations do not possess a mean nor a variance. Random observations that obey a fractal observation are characterized by having smaller values than lesser values repeated at different scales. Such fractal distributions are represented by the power-law probability density function:

$$f(x) = \frac{\lambda - 1}{\theta} \left(\frac{x}{\theta}\right)^{-\lambda}$$

Where, λ is mean fractal dimension

Ө is minimum value of the data and

For calculating probability, we need cumulative distributive function (C.D.F.) which is:

$$F(x) = \int_\theta^x f(x)\, dx = 1 - \left(\frac{x}{\theta}\right)^{1-\lambda}$$

And, $P(X_k < X < X_{k+}) = F_{x_{k+1}} - Fx_k$ for calculating probabilities in different ranges for magnitudes.

**Test Algorithm:**

- Sorting $x_1, x_2, \ldots, x_n$ from smallest to highest. Denote the sorted values by x(1) ≤ x(2) ≤… x (n) and assigning ranks 1,2…n.

- Assigning ranks(α) to each row of the data by using the following formula:

$$\alpha = \frac{Rank(x_i)}{\theta}$$

The rank function assigns an order to each observation $x_i$, and dividing by n normalizes this rank, giving a proportion that ranges between 0 and 1.

Essentially, ∝ shows the relative position of the data point $x_i$ within the entire dataset.

- Calculating Fractal dimension (λ) for each row of the data, calculated by:

$$\lambda = 1 - \frac{\log(1-\alpha)}{\log\left(\frac{x_i}{\theta}\right)}$$

This formula calculates the **fractal dimension (λ)**, which captures the complexity or irregularity of the data over different scales.

- Calculating mean fraction dimension $(\hat{\lambda}) = \frac{1}{n-1}\sum_{k=1}^{n-1} \lambda_k$

Similarly for calculating multifractal probability:

- $g(\lambda) = Ae^{-k\lambda}$, λ>1, where $k = \frac{1}{1-\bar{\lambda}}$ and $A = ke^k$

  g(y) is a prior distribution λ for solving in Bayes theorem also known as exponential decay function.

- $f(x,\lambda) = f(x/\lambda)g(\lambda) = \left(\frac{x}{\theta}\right)^{-\bar{\lambda}} ke^{k(\lambda-1)}$ this is the final formula for multifractal analysis consists of two parts probability density function f(x/λ) and exponential decay function g(y).

  f(x,λ) can represent the probability density of earthquake magnitudes or intensity at different scales (λ). Here, λ reflects various observation levels, and x represents earthquake magnitude. The exponential decay term implies that stronger seismic events are less probable, consistent with the rarity of high-magnitude earthquakes.

# 4. RESULTS:

### ❖ *Data Distribution Analysis:*

| Histogram | Q-Q Plot |
|---|---|
| 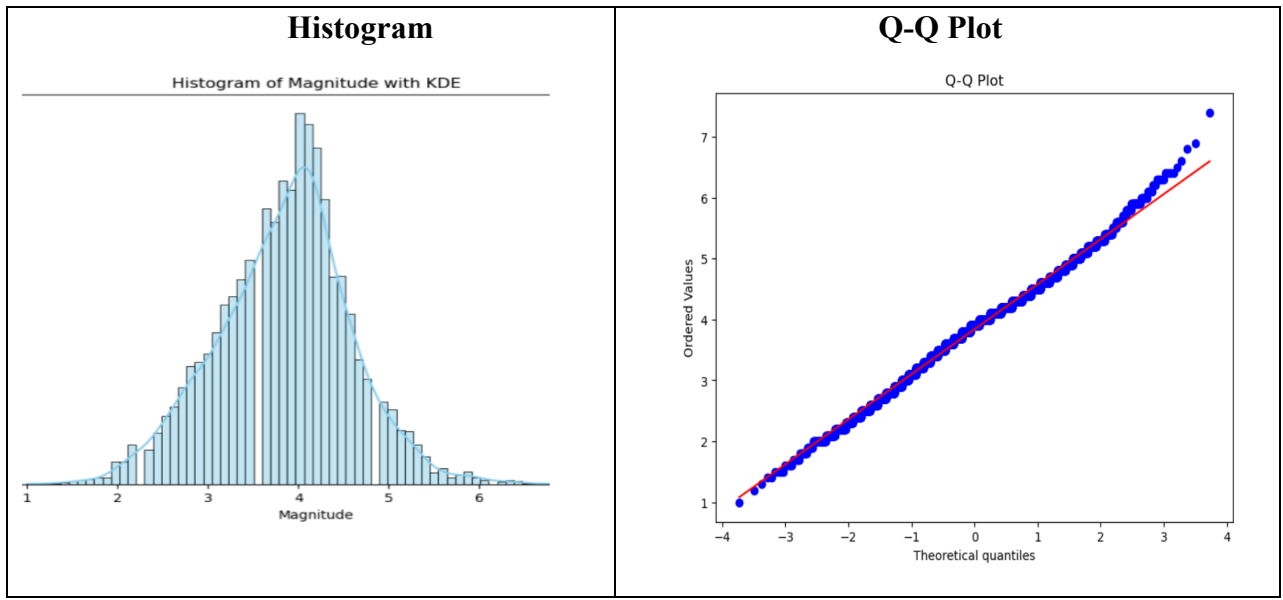 | |

Tested for normality using Kolmogorov Smirnov Test gives that Null hypothesis is rejected, so the data don't follow Normal distribution.

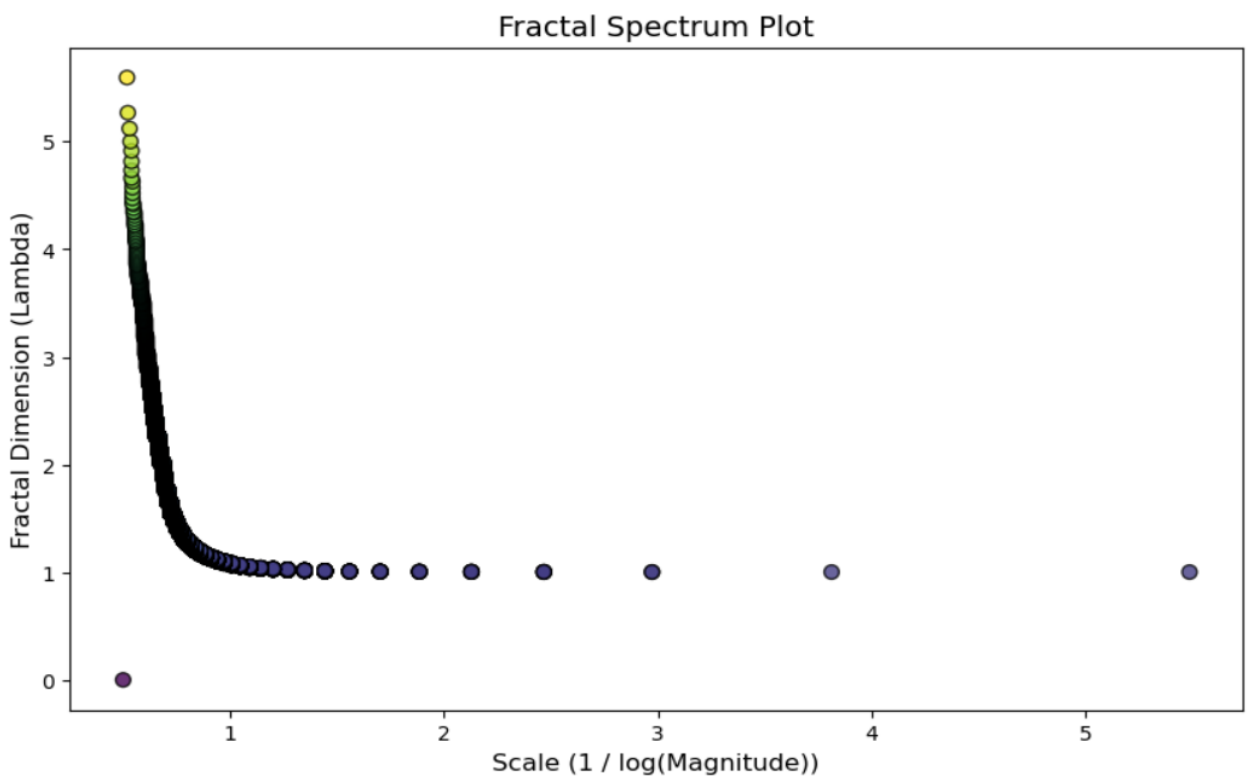

**Fig :** Plot of fractal dimensions versus scale for earthquake dimensions (Fractal Spectrum).

Calculated $\alpha$, $\lambda$ for each of the row in the dataset and $\hat{\lambda}$ mean of Fractal dimension is 1.67

Fractal Model for Earthquake Magnitudes in the India:

$$f(x) = \frac{\lambda-1}{\theta}\left(\frac{x}{\theta}\right)^{-\lambda} = \frac{0\cdot 67}{1\cdot 0}\left(\frac{x}{1\cdot 0}\right)^{-1\cdot 67}, x > 1.0$$

Using this model, we can estimate the probabilities of occurrence of earthquakes of various magnitudes in the future.

Table 1. Earthquake magnitudes in the India (mono-fractal):

| Range Magnitude of Earthquake (Richter Scale) | Probability of Occurrence | Probability in Percent | Earthquake effect |
|---|---|---|---|
| 1.0 to 1.9 | 0.3518521762001471 | 35.19 | Micro |
| 2.0 to 2.9 | 0.13898773454273844 | 13.90 | Minor I |
| 3.0 to 3.9 | 0.0773267095503164 | 7.73 | Minor II |
| 4.0 to 4.9 | 0.05021981577366985 | 5.02 | Light |
| 5.0 to 5.9 | 0.03566530101589832 | 3.56 | Moderate |
| 6.0 to 6.9 | 0.026854506572401016 | 2.68 | Strong |
| 7.0 to 7.9 | 0.02107332974701706 | 2.10 | Major |
| 8.0 to 8.9 | 0.01705346496107174 | 1.70 | Great |
| 9.0 to 10.0 | 0.015570985199882148 | 1.55 | Greatest |

Table 2. Earthquake magnitudes in the India (multi-fractal):

| Range Magnitude of Earthquake (Richter Scale) | Probability of Occurrence | Probability in Percent | Earthquake effect |
|---|---|---|---|
| 1.0 to 1.9 | 0.32957 | 32.95 | Micro |
| 2.0 to 2.9 | 0.13019 | 13.02 | Minor I |
| 3.0 to 3.9 | 0.07243 | 7.24 | Minor II |
| 4.0 to 4.9 | 0.04704 | 4.70 | Light |
| 5.0 to 5.9 | 0.03341 | 3.34 | Moderate |
| 6.0 to 6.9 | 0.02515 | 2.52 | Strong |
| 7.0 to 7.9 | 0.01974 | 1.97 | Major |
| 8.0 to 8.9 | 0.01597 | 1.60 | Great |
| 9.0 to 10.0 | 0.01459 | 1.46 | Greatest |

# 5. DISCUSSION:

*Comparison and Insights*: The comparison between the mono-fractal and multifractal models of earthquake magnitudes in India reveals significant insights. The mono-fractal model tends to overestimate the probabilities for lower magnitudes and may not fully capture the complexity of seismic data at larger scales. In contrast, the multifractal model, which accounts for varying scaling properties, offers a more precise estimation, particularly for mid-to-high magnitude ranges.

*Implications of Monofractal Results:* The monofractal approach is suitable for a basic understanding of earthquake distribution and can provide a straightforward analysis. The high probability of occurrence at lower magnitudes (e.g., 35.19% for 1.0 to 1.9) suggests that monofractal models emphasize the prevalence of smaller seismic events. However, this model may fail to capture the detailed variability found in more complex seismic behaviours, leading to potential oversimplification.

*Implications of Multifractal Results:* The multifractal model, with its slightly lower probability for lower magnitudes (e.g., 32.95% for 1.0 to 1.9) and more variable estimates for larger magnitudes, suggests a more nuanced analysis. This model better reflects the natural complexity and variability in seismic data, capturing the heterogeneous distribution of earthquakes. For higher magnitudes, the multifractal approach shows a decrease in probabilities compared to the monofractal, which can indicate its effectiveness in representing the rarity and scale variability of significant seismic events.

*Interpretation of Earthquake Risk:* The multifractal model provides a clearer picture of the distribution and likelihood of different magnitude earthquakes, making it a valuable tool for earthquake risk assessment and disaster management. The lower probabilities for higher magnitudes indicate that while these events are rare, their risk is more accurately represented through the multifractal approach. This has important implications for regions that need precise modelling for emergency preparedness and infrastructure resilience.

*Advantages of Multifractal Analysis:* The multifractal analysis accounts for varying scales within the data and provides a detailed examination that helps in understanding the heterogeneity in earthquake magnitudes. It suggests that earthquake data cannot be represented by a single fractal dimension but rather a spectrum of dimensions that capture local irregularities and scale variability.

*Limitations and Future Work:* While the multifractal model offers more detailed insights, it requires a more complex analysis and computational resources. Future research could focus on applying multifractal models to longer data periods or integrating them with other statistical tools to enhance earthquake prediction accuracy. Additionally, testing these models on different regional seismic data sets can validate their effectiveness and improve understanding.

*Practical Implications:* For policymakers and disaster management authorities, the findings highlight the importance of adopting multifractal approaches for more accurate risk modelling. This can improve the allocation of resources for mitigation strategies, public safety measures, and urban planning in seismically active regions

# 6. CONCLUSION:

The analysis of earthquake magnitude data using monofractal and multifractal models has provided meaningful insights into the distribution and probability of seismic events in India. The monofractal approach, while simple and straightforward, tends to generalize the data and may not adequately capture the complex nature of seismic activity, especially for mid-to-high magnitudes. On the other hand, the multifractal model provides a more refined and realistic representation, capturing the scale variability and heterogeneity of earthquake data.

The results show that the multifractal model is better suited for understanding and modelling the true distribution of earthquake magnitudes, as it accounts for the inherent complexity and scaling behaviours within the data. This has significant implications for seismic risk assessment and disaster preparedness, as it allows for more accurate forecasting of the probability and impact of future seismic events.

In conclusion, while both models have their merits, the multifractal approach demonstrates superior performance in representing the distribution of earthquake magnitudes. Future work should explore integrating multifractal analysis with other advanced statistical and machine learning techniques to enhance the predictive power and reliability of earthquake risk models. This will ultimately contribute to more informed decision-making and effective risk management strategies for seismic-prone regions.